\documentclass[aps,prl,twocolumn,superscriptaddress,showpacs]{revtex4}

\usepackage{graphicx}
\begin{document}
\topmargin -0.5in

\title{Density Induced Interchange of Anisotropy Axes at Half-Filled High Landau Levels}

\author{J. Zhu}
\affiliation{Department of Physics, Columbia University, New York,
New York 10027}
\author{W. Pan}
\affiliation{Department of Electrical Engineering, Princeton
University, Princeton, New Jersey 08544}
\author{H.L. Stormer}
\affiliation{Department of Physics, Columbia University, New York,
New York 10027}\affiliation{Department of Applied Physics and
Applied Mathematics, Columbia University, New York, New York
10027} \affiliation{Bell Labs, Lucent Technologies, Murray Hill,
New Jersey 07974}
\author{L.N. Pfeiffer}
\affiliation{Bell Labs, Lucent Technologies, Murray Hill, New
Jersey 07974}
\author{K.W. West}
\affiliation{Bell Labs, Lucent Technologies, Murray Hill, New
Jersey 07974}

\date{\today}

\begin{abstract}
We observe density induced 90$^{\circ}$ rotations of the
anisotropy axes in transport measurements at half-filled high
Landau levels in the two dimensional electron system, where stripe
states are proposed ($\nu$=9/2, 11/2, etc). Using a field effect
transistor, we find the transition density to be
$2.9\times10^{11}$cm$^{-2}$ at $\nu$=9/2. Hysteresis is observed
in the vicinity of the transition. We construct a phase boundary
in the filling factor-magnetic field plane in the regime
$4.4<\nu<4.6$. An in-plane magnetic field applied along either
anisotropy axis always stabilizes the low density orientation of
the stripes.
\end{abstract}
 \maketitle

The physics of electrons in higher Landau levels has recently
drawn much attention due to the discovery of large transport
anisotropies at half fillings ($\nu=9/2, 11/2$ etc) in high
quality two dimensional electron gases (2DEG) in
GaAs\cite{Lilly99,Du99}. At these filling factors, longitudinal
magneto resistances R$_{xx}$ of samples grown on (001)-oriented
GaAs show strong maxima in the $\langle1\bar{1}0\rangle$ direction
and deep minima in the $\langle110\rangle$ direction. The ratio of
the resistance values can exceed 1000:1 in high quality samples.
This large anisotropy is viewed as strong evidence for the
formation of a unidirectional Charge Density Wave or so called
``stripe state'' around these filling
factors\cite{Koulakov96,Fogler96,Moessner96,Fogler01}.
Calculations based on the Hartree-Fock approximation show the
stripes to form as the result of the competition between the
short-range attractive exchange interaction and the long-range
Coulomb repulsion. Recently there appeared calculations beyond the
Hartree-Fock
approximation\cite{Fradkin99,Fertig99,Rezayi99,MacDonald00,Shibata01}.
Some of them propose the existence of liquid crystalline states
with stripe ordering and broken rotational
symmetry\cite{Fradkin99,MacDonald00}. Employing an edge state
transport mechanism, the ``easy''(low resistance) axis is along
the stripes whereas the ``hard'' (high resistance) axis is
perpendicular to the stripes. This identification is supported by
experiments\cite{Willett01current}.

In spite of the strong evidence supporting the formation of a
stripe state, the origin of its preferred orientation remains
poorly understood\cite{Fil00,Takhtamirov01,Rosenow01}. Theory does
not \textit{a priori} provide a preferred direction for the
stripes. However, all previous experiments have identified the
$\langle110\rangle$ direction as the ``easy'' axis. It is often
assumed that anisotropic imperfections at GaAs/AlGaAs interface
act as the native symmetry breaking potential. Anisotropic
roughness has been observed in GaAs layers grown by Molecular Beam
Epitaxy\cite{MBE}. Two groups have investigated the correlation
between the surface roughness and the orientation of the stripes
which form a few thousand $\AA$ below the surface. Atomic Force
Microscope images reveal the presence of roughness elongated along
both the $\langle110\rangle$ and the $\langle1\bar{1}0\rangle$
directions. However, its correlation with the orientation of the
stripes remains controversial\cite{Willett01surface,Cooper01}.
Theoretical studies on the influence of a simple periodic
modulation on the orientation of the stripes suggest that they
prefer to align \emph{perpendicular} to a weak
potential\cite{Yoshioka01,Aoyama01}. However, a parallel alignment
may occur in the strong potential limit\cite{Aoyama01}.

Previous experiments that identified the $\langle110\rangle$
direction as the ``easy'' axis were limited to fixed densities in
the range of
$1.5-3.0\times10^{11}$cm$^{-2}$\cite{Du99,Lilly99,Cooper01,Willett01surface}.
Furthermore, the correlation between the orientation of the
stripes and the native symmetry breaking potential is difficult to
deduce since the latter is likely to vary from specimen to
specimen. Employing a tunable density Heterostructure Insulated
Gate Field Effect Transistor (HIGFET) in our experiments, we have
the unique opportunity to access a wide density regime and to
change the 2DEG density \textit{in situ} without affecting other
parameters. Our data show that as the density of the 2DEG is
raised above $2.9\times10^{11}$cm$^{-2}$, the ``easy'' axis
rotates from the $\langle110\rangle$ direction to the
$\langle1\bar{1}0\rangle$ direction. This result demonstrates that
the pinning direction of the stripes is \emph{not} unique.

\begin{figure}
\includegraphics{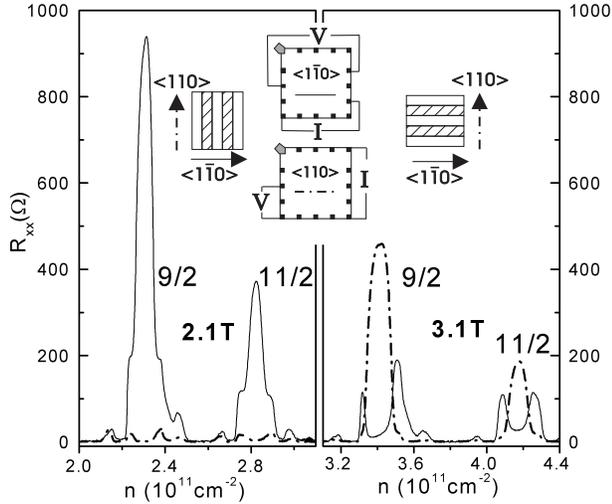}
\vspace{-0.1in} \caption{\label{density} Interchange of anisotropy
axes of the $\nu=9/2$ state with increasing densities. The two
central insets show the contact configurations used to measure the
anisotropy. Data from the $\langle1\bar{1}0\rangle$ configuration
are represented by solid lines and data from the
$\langle110\rangle$ configuration by dash-dotted lines. In the
left panel, the stripes align along the $\langle110\rangle$
direction. In the right panel, they align along the
$\langle1\bar{1}0\rangle$ direction. See insets.}
\end{figure}

Our HIGFET consists of a 600$\mu$m square mesa with edges along
the $\langle110\rangle$ and the $\langle1\bar{1}0\rangle$
directions. The structure of the specimen consists of a (001) GaAs
substrate, overgrown by MBE with 0.5$\mu$m of GaAs. A 50$\AA$
layer of AlAs is deposited, followed by a 4000$\AA$ layer of
Al$_{0.32}$Ga$_{0.68}$As and capped by a heavily doped GaAs
n$^{+}$ layer, serving as a top gate. The AlAs layer at the
interface with GaAs is intended to improve mobility. Sixteen
Ni-Ge-Au contact pads are placed evenly along the edges of the
mesa using standard optical lithography. One corner pad provides
the contact to the top gate (see inset to Fig.\ref{density}).
Biasing the gate, we are able to continuously change the 2DEG
density from $5\times10^{9}$cm$^{-2}$ to
$4.9\times10^{11}$cm$^{-2}$. The peak mobility is
$1.1\times10^7$cm$^2$/Vsec at $2.3\times10^{11}$cm$^{-2}$. The
2DEG density has a reproducible linear dependence on gate voltage.
Only the lowest electronic subband is occupied for the whole
density range. Fragile fractional quantum Hall states such as the
5/2 state and those around 3/2 are well developed for a wide range
of densities, attesting to the high quality of the sample. The
anisotropies at 9/2 and 11/2 are large, enabling us to identify
the anisotropy axes unambiguously.

The electrical measurements are carried out in a dilution
refrigerator with a base temperature of 20mK using standard low
frequency (13-21Hz) lock-in techniques. An excitation current of
10nA is used to avoid electron heating. R$_{xx}$ traces are taken
at fixed perpendicular magnetic field while sweeping gate voltage
to change density. The left and right panels of Fig.~\ref{density}
show traces taken at 2.1 Tesla and 3.1 Tesla, respectively. At
2.1T, the 9/2 state occurs at $2.3\times10^{11}$cm$^{-2}$. At
3.1T, it occurs at $3.4\times10^{11}$cm$^{-2}$. The two central
insets in Fig.~\ref{density} show the two orthogonal contact
configurations used to measure the anisotropy\cite{Contact}. They
are labelled $\langle110\rangle$ and $\langle1\bar{1}0\rangle$
according to the direction of the current flow. The
$\langle1\bar{1}0\rangle$ configuration is always represented by a
solid line and the $\langle110\rangle$ configuration by a dash
dotted line. In the left panel, at 2.1T, the ``easy'' axis is
along the $\langle110\rangle$ direction, in agreement with all
previous experiments. In the right panel, at 3.1T, the ``easy''
axis is \emph{rotated} 90$^{\circ}$ and points now along the
$\langle1\bar{1}0\rangle$ direction. We conclude that the
underlying stripe state has also rotated 90$^{\circ}$ and is now
aligned along the $\langle1\bar{1}0\rangle$ direction. This
demonstrates that the pinning direction of the stripes is
\emph{not} unique. This observation is the most striking result of
our experiments.

Our results indicate that the pinning mechanism of the stripes is
more complex than previously assumed. A few conclusions can be
drawn. First, contrary to earlier results, a
$\langle1\bar{1}0\rangle$ orientation of the stripes \textit{does}
exist in a perpendicular magnetic field. Second, assuming that the
stripes are pinned by a unidirectional periodic potential, these
stripes can be aligned either \emph{parallel} or
\emph{perpendicular} to it. Although we do not know the mechanism
responsible for the reorientation, we identify several
consequences of changing density and discuss their effects. With
increasing density, the electron wave function is squeezed and
pressed harder against the interface. Therefore, the electrons
experience the interface potential more strongly. On the other
hand, with increasing density, screening also increases,
effectively weakening the interface potential. In addition, the
period of the stripes, calculated to be a few magnetic lengths at
$\nu=9/2$ \cite{Koulakov96}, decreases with increasing density.
Any of these factors could lead to the observed reorientation of
the stripes\cite{Aoyama01}. Furthermore, we can not rule out the
possibility of the coexistence of two orthogonal periodic
potentials in our sample. Changing the period of the stripes could
shift the relative effectiveness of these two potentials and may
cause a reorientation.

\begin{figure}
\includegraphics{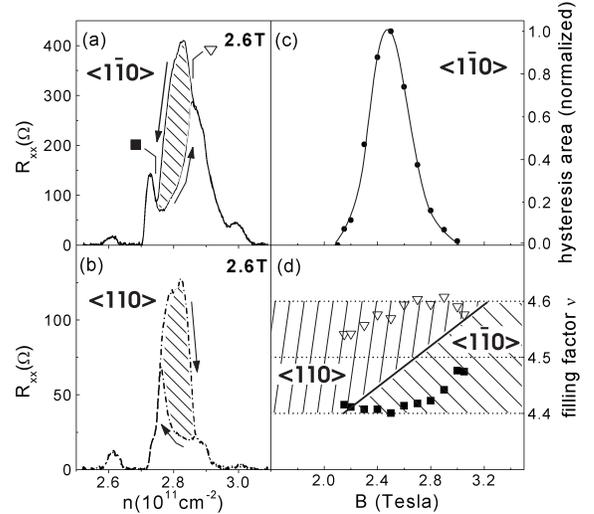}
\vspace{-0.1in}
 \caption{\label{hysteresis} Hysteresis at 2.6T in
the $\langle1\bar{1}0\rangle$ contact configuration(a) and the
$\langle110\rangle$ contact configuration(b). Sweep directions are
indicated by arrows and the hysteretic areas are hatched. Sweep
rate is $\sim1.0\times10^{10}$cm$^{-2}$/min. c) Area of the
hysteresis loop as a function of magnetic field. d) Filling factor
range of the hysteresis in the $\nu$-B plane. Symbols correspond
to (a). The solid line represents the general position of the
phase boundary.}
\end{figure}

To find the transition density for the reorientation of the
stripes, we take density sweeps at fixed magnetic fields from 2.0T
to 3.3T in steps of 0.1T and identify 2.7T and the corresponding
density n$=2.9\times10^{11}$cm$^{-2}$ as the transition point at
$\nu=9/2$. From 2.2T to 3.0T, we observe hysteresis between
density up-sweeps and down-sweeps in a narrow range of filling
factors ($\Delta\nu\sim0.2$) around 9/2. As an example,
Fig.~\ref{hysteresis}a and b show the hysteresis at 2.6T and 40mK
in both contact configurations. Sweep directions are indicated by
arrows and the hysteretic area is hatched. The hysteresis
decreases with increasing temperature and disappears altogether at
~90mK while the anisotropy remains although it is weaker. These
90mK data unambiguously identify the orientation of the stripes in
close vicinity to $\nu=9/2$. Fig.~\ref{hysteresis}c shows the area
of the hysteresis loop in the $\langle1\bar{1}0\rangle$
configuration at 40mK as a function of magnetic field. It peaks
near 2.5T and vanishes at 2.1T and 3.0T. In
Fig.~\ref{hysteresis}d, we plot the range of the hysteretic
behavior in the $\nu$-B plane. This hysteresis ``ellipse'' is
bounded by $\nu=4.4$ and $\nu=4.6$, which coincides with the
filling factor range in which the stripe state is expected to
exist theoretically\cite{Fogler96}.

In general, the presence of hysteresis indicates a first order
phase transition. As the system crosses the transition point, the
initial phase continues to survive due to some pinning mechanism.
In our sample, the phase transition corresponds to the
reorientation of the stripes while impurities act as the pinning
mechanism. In the $\nu$-B plane, the stripe state exists in a
horizonal band between $\nu$=4.4 and $\nu$=4.6. The hysteresis
``ellipse'' is embedded within this band, indicating the
transition region. To the left of the ``ellipse'', the stripes
have the $\langle110\rangle$ orientation. To the right of the
``ellipse'', the stripes have the $\langle1\bar{1}0\rangle$
orientation. Therefore, the boundary separating the two phases
must stretch through the whole bubble from the left side to the
right side. We know from the 90mK data that the boundary
intercepts the $\nu=$4.5 line at 2.7T. Furthermore, the
orientation of the stripes along the border of the hysteresis
``ellipse'' can be deduced from data such as
Fig.~\ref{hysteresis}a and b since hysteresis always prolongs the
initial phase. For example, at 2.6T, on density up-sweeps, the
stripes are along the $\langle1\bar{1}0\rangle$ direction, as seen
from the minimum in the $\langle1\bar{1}0\rangle$ configuration
and the maximum in the $\langle110\rangle$ configuration. This
indicates that the stripes near $\nu=4.4$ have the
$\langle1\bar{1}0\rangle$ orientation. Conversely, density
down-sweeps indicate the stripes near $\nu=4.6$ have the
$\langle110\rangle$ orientation. Similar identifications can be
made for most other points along the border of the hysteresis
``ellipse''. From these inputs, we conclude that the phase
boundary must run from the lower left to the upper right of the
``ellipse'', intercepting the $\nu=4.5$ line at 2.7T. Although the
exact position of the phase boundary can not be derived from our
data, a straight line as shown in Fig.~\ref{hysteresis}d indicates
its general position. Within the $4.4<\nu<4.6$ band, the stripes
align along the $\langle110\rangle$ direction to the left of the
boundary. To the right of the boundary, the stripes align along
the $\langle1\bar{1}0\rangle$ direction. This analysis provides a
phase diagram for the stripes around $\nu=9/2$. Hystereses with
much smaller areas are also observed around $\nu=11/2$.

It remains unclear why in previous studies samples with densities
inside the transition region of Fig.~\ref{hysteresis}d neither
show the $\langle1\bar{1}0\rangle$ orientation nor the hysteresis
\cite{Cooper01,Willett01surface}. It suggests the importance of
the individual sample details. Along this line, the thin AlAs
layer at the interface may play a role in establishing the
specific transition density.

The application of an in-plane magnetic field has become a useful
tool in investigating the native symmetry breaking potential since
it gives us additional control over the orientation of the
stripes. In previous fixed density samples, the stripes always
aligned perpendicular to a sufficiently strong ($\sim$0.5T)
in-plane field B$^{ip}$\cite{Pan99field,Lilly99field}. This means
that an in-plane field parallel to the B$^{ip}$=0 orientation of
the stripes reorients the stripes to the perpendicular alignment.

\begin{figure}
\includegraphics{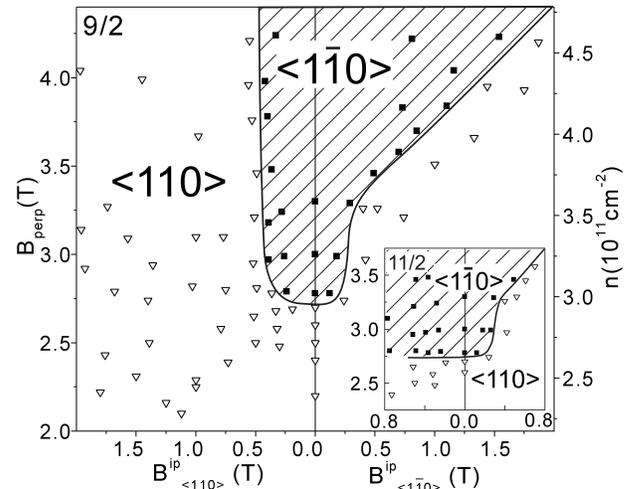}
\vspace{-0.2in} \caption{\label{diagram} Phase diagram of the
orientation of the stripes at $\nu=9/2$ in the B$^{perp}$-B$^{ip}$
plane. Solid squares represent the $\langle1\bar{1}0\rangle$
orientation of the stripes. Hollow triangles represent the
$\langle110\rangle$ orientation. In the left panel, B$^{ip}$
points along the $\langle110\rangle$ direction. In the right
panel, B$^{ip}$ points along the $\langle1\bar{1}0\rangle$
direction. The inset shows the phase diagram at $\nu=11/2$. Units
are the same as in the $\nu=9/2$ diagram.}
\end{figure}

We apply an in-plane magnetic field B$^{ip}$ to the 2DEG by
rotating the HIGFET \emph{in situ} in our dilution refrigerator.
Two cooldowns are required to align B$^{ip}$ along the
$\langle110\rangle$ or the $\langle1\bar{1}0\rangle$ direction
respectively. The rotator assembly is able to reach a base
temperature of 70mK. Tilting angles were determined from a
calibration of the rotator and from positions of well defined QHE
and FQHE minima. The accuracy is better than 1$^{\circ}$. By
sweeping density at a fixed angle, i.e. fixed perpendicular
magnetic field, B$^{perp}$, and fixed B$^{ip}$, we determine the
orientation of the stripes at $\nu=9/2$. Fig.~\ref{diagram} shows
the phase diagram in the B$^{perp}$-B$^{ip}$ plane constructed
from such data. The left panel represents data taken with B$^{ip}$
along the $\langle110\rangle$ direction. The right panel
represents data taken with B$^{ip}$ along the
$\langle1\bar{1}0\rangle$ direction. A $\langle110\rangle$
orientation of the stripes is represented by a hollow triangle and
a $\langle1\bar{1}0\rangle$ orientation by a solid square.

In order to assess the properties of the phase diagram, we first
focus on the B$^{ip}$=0 axis which simply reflects the
reorientation of the stripes discussed earlier: below
B$^{perp}$=2.7T (equivalent to n$=2.9\times10^{11}$cm$^{-2}$,
right axis), the stripes have the $\langle110\rangle$ orientation
whereas above 2.7T, the $\langle1\bar{1}0\rangle$ orientation
prevails. As we increase B$^{ip}$ along either the
$\langle110\rangle$ or the $\langle1\bar{1}0\rangle$ direction,
the $\langle110\rangle$ to $\langle1\bar{1}0\rangle$ transition
moves to higher densities. The density shift indicates a B$^{ip}$
induced energy gain of the $\langle110\rangle$ phase. It can be
viewed as a sum of two terms: an \emph{isotropic} term and an
\emph{anisotropic} term. The isotropic term is independent of the
direction of B$^{ip}$. The anisotropic term reflects the
difference in energy gain between a B$^{ip}$ parallel to the
stripes and a B$^{ip}$ perpendicular to the stripes. Two
conclusions can be drawn from the phase diagram. First, the
roughly symmetric shape of the phase boundary indicates that in
our sample, the dominant in-plane magnetic field effect is an
isotropic energy gain of the $\langle110\rangle$ phase. The
anisotropic term only becomes significant for
$\left|B^{ip}\right|>0.4T$. Second, in the high B$^{ip}$ regime,
in the right panel, the transition to the
$\langle1\bar{1}0\rangle$ phase occurs at lower densities than in
the left panel. From this, we conclude that stripes oriented
\emph{parallel} to B$^{ip}$ are energetically favored as compared
to stripes oriented perpendicular to B$^{ip}$. Both conclusions
are in contrast to previous fixed density samples where the
\emph{anisotropic} term, which favors a \emph{perpendicular}
alignment of the stripes with respect to B$^{ip}$,
dominates\cite{Pan99field,
Lilly99field,Jungwirth99well,Stanescu00,Twosubbandnote}. Together
with previous results, we must conclude that the reaction of the
stripes to an in-plane magnetic field is non-universal and
sensitively depends on the details of individual samples. The
unique density tunability of a HIGFET is essential to such a
systematic study. In fact, diverse behaviors are seen even within
one specimen. As an inset to Fig.~\ref{diagram}, we show the phase
diagram of the 11/2 state, constructed during the same cooldowns
as the 9/2 state using the same method. The B$^{ip}$=0 transition
also occurs at$\sim$2.7T. We can clearly see that the isotropic
behavior observed at $\nu$=9/2 is missing. The differences between
the behaviors of the 9/2 state and the 11/2 state strongly attest
to the subtlety of the interactions involved. Examining other
half-fillings, we note that the isotropic states of $\nu=5/2$ and
$\nu=7/2$ become anisotropic in an in-plane field and the ``hard''
axis is always along the in-plane field direction. This is
consistent with previous experimental findings and theoretical
calculations\cite{Pan99field,Lilly99field,Jungwirth99well}.

In summary, we have observed a density induced reorientation of
the stripe state at $\nu=9/2$. As the density of the 2DEG is
raised above $2.9\times10^{11}$cm$^{-2}$, the stripes rotate from
the $\langle110\rangle$ orientation to the
$\langle1\bar{1}0\rangle$ orientation. Our results demonstrate
that the pinning direction of the stripes is \textit{not} unique.

\begin{acknowledgments}
We thank A. Millis and S. Simon for helpful discussions and K.
Baldwin and E. Chaban for experimental assistances. Financial
support from the W. M. Keck Foundation is gratefully acknowledged.
\end{acknowledgments}

%\bibliography{anisotropy_final}

\begin{thebibliography}{28}
\bibitem[{Con()}]{Contact}
\bibinfo{note}{Current and voltage contacts in the $\langle1\bar{1}0\rangle$
  configuration do not align exactly along the $\langle1\bar{1}0\rangle$
  direction (14$^{\circ}$ off). Such a small misalignment affects the
  observations of anisotropy very little due to the current channeling property
  of the van der Pauw configuration\cite{Simon99,Willett01current}.}
\bibitem[{Two()}]{Twosubbandnote}
\bibinfo{note}{The only exception is in a square well sample with two occupied
  electronic subbands where stripes parallel to B$^{ip}$ are found at
  intermediate field values\cite{Pan00subband}.}
\end{thebibliography}
\end{document}